\documentclass[aps, prl, amsmath, floats,floatfix, twocolumn, superscriptaddress, 
showpacs]{revtex4}

\usepackage{amssymb}
\usepackage{amsmath}
\usepackage{verbatim}
\usepackage{mathrsfs}
\usepackage{amsfonts}
\usepackage{latexsym}
\usepackage{epsfig}
\usepackage{color}
\usepackage{graphicx,subfigure}
\usepackage{units}
\usepackage{slashbox}

\begin{document}


\definecolor{orange}{rgb}{0.9,0.45,0}

\newcommand{\re}{\mbox{Re}}
\newcommand{\im}{\mbox{Im}}

\renewcommand{\t}{\times}

\long\def\symbolfootnote[#1]#2{\begingroup%
\def\thefootnote{\fnsymbol{footnote}}\footnote[#1]{#2}\endgroup}


\title{Explosion and final state of an unstable Reissner-Nordstr\"om black hole} 

\author{Nicolas Sanchis-Gual}
\affiliation{Departamento de
  Astronom\'{\i}a y Astrof\'{\i}sica, Universitat de Val\`encia,
  Dr. Moliner 50, 46100, Burjassot (Val\`encia), Spain}

\author{Juan Carlos Degollado} 
\affiliation{
Instituto de Ciencias F\'isicas, Universidad Nacional Aut\'onoma de M\'exico,
Apdo. Postal 48-3, 62251, Cuernavaca, Morelos, M\'exico.}
\affiliation{
Departamento de Ciencias Computacionales,
Centro Universitario de Ciencias Exactas e Ingenier\'ia, Universidad de Guadalajara\\
Av. Revoluci\'on 1500, Colonia Ol\'impica C.P. 44430, Guadalajara, Jalisco, Mexico}

\author{Pedro J. Montero} 
\affiliation{Max-Planck-Institut f{\"u}r Astrophysik, Karl-Schwarzschild-Str. 1, 85748, Garching 
bei M{\"u}nchen, Germany}

\author{Jos\'e A. Font}
\affiliation{Departamento de
  Astronom\'{\i}a y Astrof\'{\i}sica, Universitat de Val\`encia,
  Dr. Moliner 50, 46100, Burjassot (Val\`encia), Spain}
\affiliation{Observatori Astron\`omic, Universitat de Val\`encia, C/ Catedr\'atico 
  Jos\'e Beltr\'an 2, 46980, Paterna (Val\`encia), Spain}

\author{Carlos Herdeiro}
\affiliation{Departamento de F\'{\i}sica da Universidade de Aveiro and CIDMA, Campus de Santiago, 
3810-183 Aveiro, Portugal}


\date{December 2015}


\begin{abstract} 
A Reissner-Nordstr\"om black hole (BH) is superradiantly unstable against spherical perturbations of a charged scalar field, enclosed in a cavity, with frequency lower than a critical value. We use numerical relativity techniques to  follow the development of this unstable system -- dubbed \textit{a charged BH bomb} -- into the non-linear regime, solving the full Einstein--Maxwell--Klein-Gordon equations, in spherical symmetry. We show that: $i)$ the process stops before all the charge is extracted from the BH; $ii)$ the system settles down into a hairy BH: a charged horizon in equilibrium with a scalar field condensate, whose phase is oscillating at the (final) critical frequency. For low scalar field charge, $q$, the final state is approached smoothly and monotonically. For large $q$, however,  the energy extraction overshoots and an explosive phenomenon, akin to a \textit{bosenova}, pushes some energy back into the BH. The charge extraction, by contrast, does not reverse. 
\end{abstract}




\maketitle


{\bf {\em Introduction.}} A remarkable feature of rotating (Kerr) black holes (BHs) is that they may, classically, give away energy and angular momentum. A bosonic field can be the extraction mediator. Its waves, with sufficiently slowly rotating phases, are amplified when scattering off a co-rotating BH~\cite{Bardeen:1972fi,Starobinsky:1973a,Press:1972zz,Damour:76,Zouros:1979iw,Detweiler:1980uk,Cardoso:2004nk,Dolan:2007mj,Rosa:2009ei}. Trapping these \textit{superradiantly scattered} waves around the BH, the bosonic field piles up, exponentially, into a gravitating, macroscopic, Bose--Einstein-type condensate. It has been conjectured that an explosive phenomenon ensues, dubbed \textit{a BH bomb}~\cite{Press:1972zz}. Understanding the explosion and final state of the BH bomb has been an open issue  since the 1970s~\cite{Brito:2015oca}.

The BH bomb proposal was based on linear studies of the superradiant instability. The conjectured explosive regime, however, is non-linear and numerical evolutions using the full Einstein equations are mandatory to probe it. Unfortunately, the growth rates of superradiant instabilities for rotating BHs are too small~ \cite{Cardoso:2004nk,Dolan:2013}, rendering the numerical evolution of the rotating BH bomb a \textit{tour de force} with current numerical relativity (NR) technology~\cite{Okawa:2014nda,East:2014prd}.  But suggestive progress has come from two other types of non-linear studies. 
First, considering a test bosonic field with non-linear dynamics, on the Kerr BH~\cite{Yoshino:2012kn,Yoshino:2015nsa}, produced evidence that an explosive event indeed occurs, akin to the \textit{bosenova} observed in condensed matter systems~\cite{Cornish:2000zz}. 
Second, \textit{hairy BH} solutions, with a stationary geometry, of the fully non-linear Einstein-bosonic field system were found, precisely at the threshold of the instability~\cite{Herdeiro:2014goa,Herdeiro:2015gia}.

In the absence of the NR technology to address the rotating BH bomb, we are led to the more favorable situation that occurs for charged (Reissner-Nordstr\"om) BHs. An analogue process to superradiant scattering can take place, by which Coulomb energy and charge are extracted from the BH by a \textit{charged} bosonic field~ \cite{Bekenstein:1973ur,Hod:2013eea}. This occurs for sufficiently small frequency waves, and for a field with the same charge (sign) as the BH. Introducing a trapping mechanism, a \textit{charged BH bomb} forms. 
On the one hand, linear studies show that the growth rates of such charged superradiant instability can be much larger than for their rotating counterparts~\cite{Herdeiro:2013pia,Hod:2013fvl,Degollado:2013bha}. On the other hand, the instability can occur within spherical symmetry, in contrast with the rotating case that breaks even axial symmetry. These features make the study of the charged BH bomb amenable with current NR techniques. 

In this \textit{Letter}, we report NR simulations, using the full Einstein equations, of the charged BH bomb. As a simple model, we take a charged scalar field (SF) as the bosonic mediator and enclose the BH-SF system in a cavity, as a trapping mechanism. We find that the non-linear regime may be, albeit needs not be, explosive. Moreover, we establish that, regardless of how explosive the non-linear regime is, the generic final state is a \textit{hairy} BH: a charged horizon, surrounded by a SF condensate storing part of the charge and energy of the initial BH, and with a phase oscillating at the threshold frequency of the superradiant instability. Hairy BHs of this sort have been recently constructed and shown to be stable~\cite{Dolan:2015dha}. 

{\bf {\em Framework.}} We consider the Einstein-Maxwell-Klein-Gordon (EMKG) system, described by the action $\mathcal{S}=\int d^4x \sqrt{-g}\mathcal{L}$, with Lagrangian density:
\begin{equation}
\mathcal{L}=\frac{R-F_{\alpha\beta}F^{\alpha\beta}}{16\pi}-\frac{1}{2}D_\alpha \Phi (D^\alpha\Phi)^*-\frac{\mu^2}{2}|\Phi|^2 \ ,
\label{model}
\end{equation}
where $R$ is the Ricci scalar, $F_{\alpha \beta}\equiv \nabla_{\alpha}A_{\beta} - \nabla_{\beta}A_{\alpha}$, 
$A_{\alpha}$ is the electromagnetic potential, $D_\alpha$ is the gauge covariant derivative, $D_\alpha\equiv \nabla_\alpha -iqA_\alpha$, and $q$ and $\mu$ are the charge and the mass of 
the scalar field. Newton's constant, the speed of light and $4\pi\epsilon_0$ are set to one in our units.

To address, numerically, the EMKG system, we use a generalized
BSSN formulation \cite{Baumgarte98,Shibata95}, adapted to spherical symmetry~\cite{Brown:2009,Alcubierre:2010is,Montero:2012yr}, and the code described in~\cite{Sanchis-Gual:2015bh,Sanchis-Gual:2015sms}. This code was upgraded to account for
Maxwell's equations and energy-momentum tensor. The 3+1 metric split reads $ds^2=-(\alpha^2+\beta^r \beta_r)dt^2+2\beta_r dtdr+e^{4\chi}\left[a\, dr^2+ b\, r^2 d\Omega_2\right]$, where the lapse $\alpha$, shift component $\beta^r$, and the (spatial) metric functions, $\chi,a,b$ depend on $t,r$. The electric field $E^\mu=F^{\mu\nu} n_\nu$ has only a radial component and the magnetic field $B^\mu=\star F^{\mu\nu} n_\nu$ vanishes, where $n^\mu$ is the 4-velocity of the Eulerian observer~\cite{Torres:2014fga}. Spherical symmetry implies we only have to consider the equations for the electric potential, $^{(3)}\varphi=-A^\mu n_\mu$, and the 
radial component of both the vector potential, $A^{r}$, and the electric field, $E^{r}$.

At $r=r_{\rm m}$ (mirror) and beyond, the SF, $\Phi$, is required to vanish. This leads to a discontinuity in the $\Phi$ derivatives. In our scheme, however, the consequent constraint violation does not propagate towards $r<r_{\rm m}$. We further impose parity boundary conditions at the origin (puncture) for the SF.

 {\bf {\em Initial data and parameters.}} 
  The EMKG system admits as a solution the Reissner-Nordstr\"om BH with ADM mass $M$, and charge $Q$, together with a vanishing SF. We take the initial data to describe one such BH with $M=1$ and $Q=0.9$. The former will set the main scale in the problem.
Perturbing such a BH with a spherical scalar wave $\Phi=e^{-iwt}f(r)$ yields a superradiant instability if: $i)$ $w<w_{\rm c} \equiv q\phi_H$, where $\phi_H$ is the electric potential at the horizon and $ii)$ the perturbation is trapped by imposing reflecting boundary conditions for the SF at the spherical surface $r=r_{\rm{m}}$, (sufficiently) outside the horizon. 

To trigger the instability we set, as the SF initial data, a  Gaussian 
distribution of the form $ \Phi=A_0e^{-(r-r_0)^2/\lambda^2}$, with  $A_0=3\times 10^{-4}$, $r_0=7M$ and $\lambda=\sqrt{2}$ and set the mirror at $r_{\rm{m}}=14.2M$. The SF mass is fixed to $ \mu=0.1/M$ and we focus on models with different values of the SF charge $qM$, namely $qM=0.8, 5, 20$ and $40$. 

The logarithmic numerical grid extends from the origin to $r=10^4M$ and uses a maximum resolution of
$\Delta r=0.025M$. Simulations with varying resolutions have shown the expected second order convergence of the code.  An analysis of constraint violations, which we have observed to be
always around $10^{-5}$ outside
the horizon and converging away at the expected second-order rate,
together with a broader survey
of the parameter space is presented as supplemental material~\footnote{See Supplemental Material, which includes Refs.~\cite{Dolan:2015dha,Sanchis-Gual:2014nha}.}.

{\bf {\em Physical quantities.}} The extraction of energy and charge from the BH by the superradiant instability is compatible with the second law of thermodynamics. This can be checked by monitoring the irreducible mass \cite{Christodoulou:1970wf} of the BH, computed in terms of the apparent horizon (AH) area $A_{\rm{AH}}$, on each time slice, as $M_{\rm{irr}} = \sqrt{{A_{\rm{AH}}}/(16\pi)}$. For the initial RN BH, $M_{\rm irr}^{\rm ini}\simeq 0.718M$, and we will see that the final BH has a larger $M_{\rm irr}$, for all cases.

The energy transfer from the BH to the SF can be established by computing the energy stored in the latter. This is given by the (spatial) volume integral
\begin{equation}
E_{\rm{SF}}=\int^{r_{\rm m}}_{r_{\rm{AH}}}\mathcal{E}^{\rm{SF}}dV,
\label{eq:ESF}
\end{equation}
where $\mathcal{E}^{\rm{SF}}$ is the 
projection of the stress--energy tensor of the scalar field along the normal direction to the $t= $constant surfaces~\cite{Alcubierre08a}.

The charge transfer, on the other hand, is monitored by tracking both the SF charge, using a formula similar to~(\ref{eq:ESF}) replacing $\mathcal{E}^{\rm{SF}}$ by the charge density, and  the BH charge, $Q_{\rm{BH}}$, evaluated at the AH as~\cite{Torres:2014fga}
\begin{equation}
Q_{\rm{BH}}=\left(r^{2}e^{6\chi}\sqrt{ab^{2}}E^{r}\right)\big|_{\rm AH} \ .
\end{equation}

Finally, to establish the nature of the final BH, we compute the electric potential at the AH and the corresponding critical frequency, $w_{\rm c}=q\phi_H$, as
$\phi_H =  
\alpha \,^{(3)}\varphi-\beta^{r}a_{r}|_{r=r_{\rm{AH}}}$,
where $a_{r}=\gamma_{rr}A^r$ and $\gamma_{rr}$ is the corresponding component of the spatial metric~\cite{Alcubierre:2009ij}.

\begin{figure}[h!]
\begin{center}
\subfigure{\includegraphics[width=0.48\textwidth]{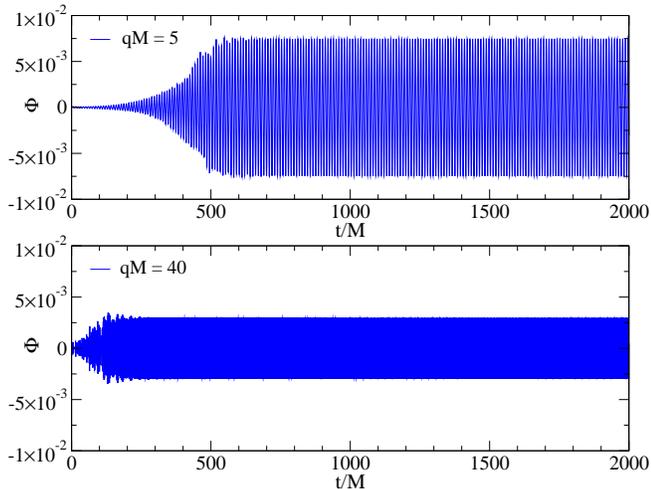}}\vspace{-0.5cm}\\
\caption{Time evolution of the SF real part, extracted at $r_{\rm{obs}}=10M$, for $qM=5$ (top) and $40$ (bottom). The imaginary part is analogous (but with opposite phase at late times).}
\label{fg:SF0}
\end{center}
\end{figure}

{\bf {\em Numerical evolutions and final state.}}
Solving numerically the EMKG system we obtain a time series for the evolution of the SF real and imaginary parts, at a chosen observation point, say, $r_{\rm{obs}}=10M$. This is illustrated in Fig.~\ref{fg:SF0} for two values of $qM$.

 Fig.~\ref{fg:SF0} demonstrates the existence of two distinct phases in the SF evolution. The first phase  is the \textit{superradiant growth phase}, known from linear theory. During this phase, the SF is amplified,  extracting energy and charge from the BH, and its amplitude grows exponentially, $|\Phi|\sim e^{t/\tau}$; a numerical fit for the e-folding time, $\tau$, is reported in~Table~\ref{tab:mod1}. The second phase, however, is outside the scope of linear/test field theory. It is the \textit{saturation and equilibrium phase}: superradiant extraction stalls at $t/M\sim 500$ ($\sim 100$) for $qM=5$ (40), and the amplification stops.  Then, after a more or less tumultuous period -- to be addressed below -- the SF amplitude remains constant for arbitrarily long evolution times. An equilibrium state between the SF and the BH is reached. 

To establish the nature of this equilibrium state, we perform a Fast Fourier Transform to obtain the oscillating frequency spectrum. The angular frequency, $\omega^{\rm fin}_{\rm SF}$, for the \textit{single} mode of oscillation in the final SF condensate, is $M\omega^{\rm fin}_{\rm SF}=0.642$ ($3.130$) for $qM=5$ (40). Then, computing the critical frequency $\omega^{\rm fin}_{\rm c}$, from the horizon electric potential of the final BH, we obtain \textit{precisely} the same value -- see~Table~\ref{tab:mod1}. Thus, these configurations are \textit{hairy} BHs that exist at the threshold of the superradiant instability. 

\begin{figure}[h!]
\begin{center}
\ \ \ \includegraphics[width=0.46\textwidth]{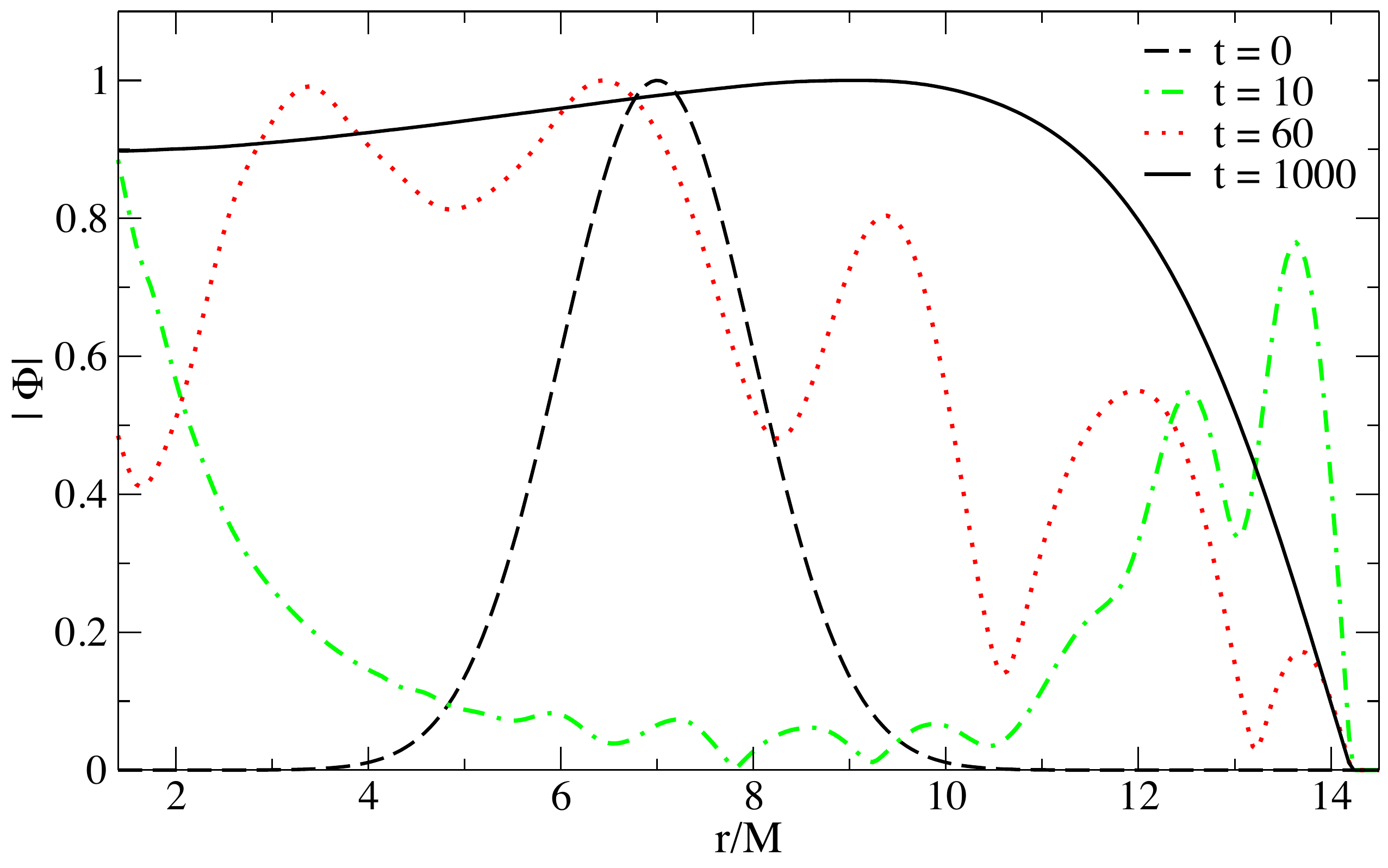}\\
\includegraphics[width=0.48\textwidth, height=0.171\textheight]{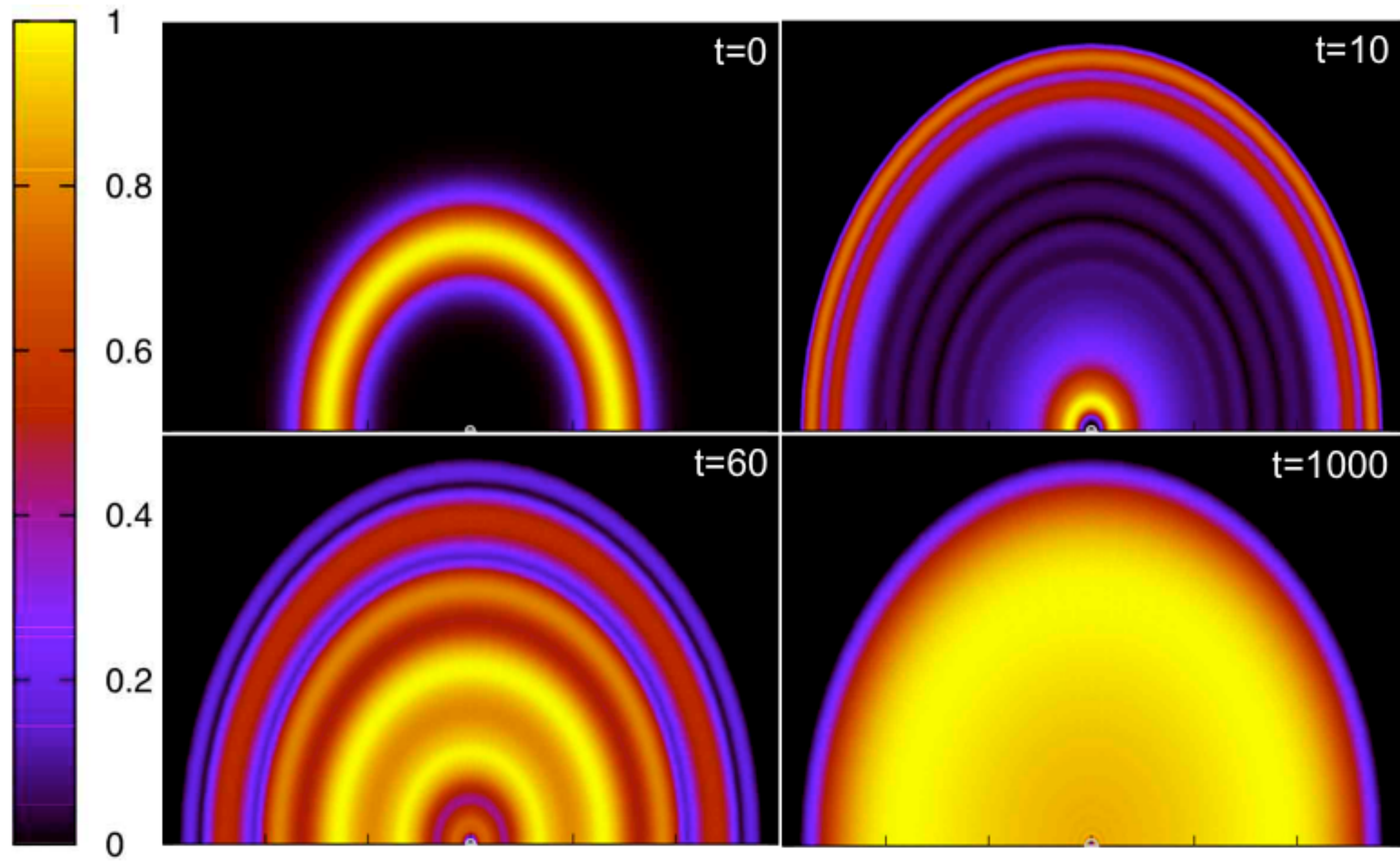}
\caption{1D (top panel) and 2D (bottom panels) snapshots of the normalized SF radial profile for $qM=40$ at times: $t/M=0, 10, 60, 1000$. The small white circles near the origin in the 2D panels mark the AH.}
\label{radialprofile}
\end{center}
\end{figure}

Charged hairy BHs in a cavity at the threshold of the superradiant instability have been recently constructed by Dolan et al.~\cite{Dolan:2015dha}, for the model~\eqref{model} with $\mu=0$. Therein it was established the existence of different families of such hairy BHs, with different numbers of nodes, $N$, for the SF amplitude between the horizon and the mirror. But only the solutions with $N=0$ are stable against perturbations.  In Fig.~\ref{radialprofile} we exhibit snapshots of the SF amplitude radial profile, at different time steps, for $qM=40$. It can be observed that whereas during the evolution the scalar amplitude exhibits several maxima and minima (and nodes exist), the final configuration has no nodes. A qualitative difference between the final state hairy BHs presented here and the stationary solutions in~\cite{Dolan:2015dha} is that the radial profiles here have a local maximum between the horizon and the mirror, which is due to the non-zero mass term.  Indeed, simulations with $\mu=0$ show no such maximum ($cf.$ supplemental material). Nevertheless, the evolutions presented here, together with the results in~\cite{Dolan:2015dha}, establish that the hairy BHs \textit{dynamically} obtained in this work are stable configurations.


\begin{figure}[h!]
\begin{center}
\subfigure{\includegraphics[width=0.48\textwidth]{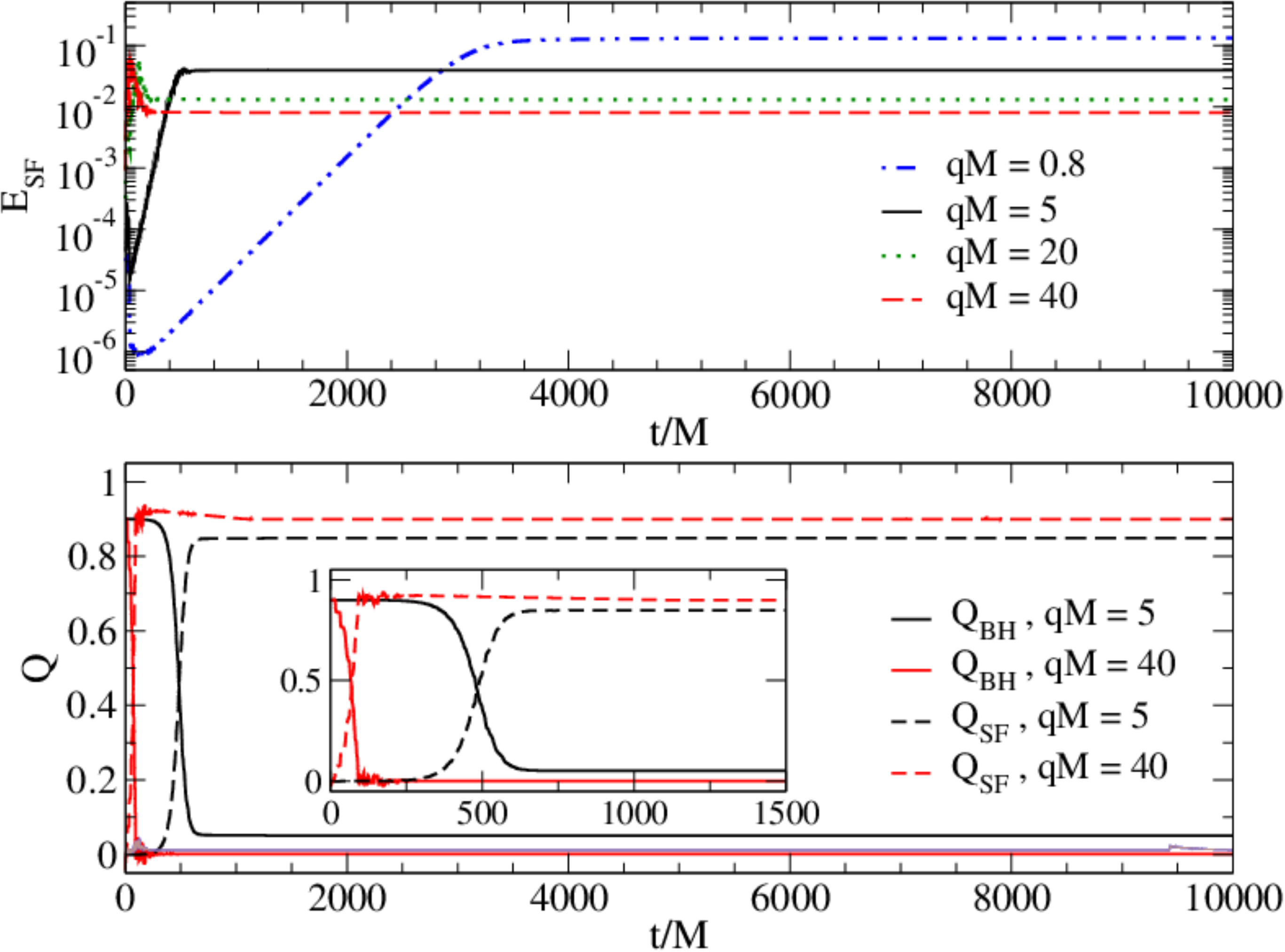}}\vspace{-0.5cm}\\
\caption{Top panel: Time evolution of the SF energy, displayed in logarithmic scale. Bottom panel: Time evolution of the charge, for both the SF and the BH. The inset zooms in the early phase of the evolution, for clarity.}
\label{fg:SF2}
\end{center}
\end{figure}

{\bf {\em Charge and energy extraction.}} We now consider in more detail the energy and charge transfer from the initial BH to the SF.  The second column in Table~\ref{tab:mod1} shows that the e-folding time of the instability, during the growth phase, decreases with increasing $qM$. This is in agreement with what can be observed in the top panel of Fig.~\ref{fg:SF2}, exhibiting the time evolution of the SF energy: comparing the curves for $qM=0.8$ and $5$, during the superradiant growth phase, the slope is larger for larger $qM$. For both these cases, the SF energy increase is essentially monotonic, until the saturation and equilibrium phase is reached. Also, one observes that the final SF energy is larger for \textit{smaller} $qM$. The corresponding quantitative values are given in the sixth column of Table~\ref{tab:mod1}. Considering that the initial perturbation has larger energy for large $qM$, $cf.$ the fifth column of Table~\ref{tab:mod1}, the ratio between the final to initial SF energy varies from $\sim 4.4\times 10^3$ to $\sim 9.0$, when $qM$ increases from $qM=0.8$ to $40$. Thus energy extraction is more efficient for lower charge coupling, corresponding to a longer and smoother superradiant growth.

\begin{table*}
\caption{Summary of physical quantities for the runs with different $qM$ (first column): e-folding time during the growth phase (second column); final oscillation frequency of the SF phase and final critical frequency (third and fourth columns); initial and final SF energy, and their ratio (fifth to seventh columns); final BH irreducible mass and ratio of the final to initial BH and SF charge (eighth to tenth columns).}
\label{tab:mod1}
\begin{ruledtabular}
\begin{tabular}{c||c|cc|ccc|ccc}
$qM$& $\tau/M$&{$\,\,M\omega^{\rm fin}_{\rm SF}$} &$M\omega_{\rm c}^{\rm fin}$&$E_{\rm SF}^{\rm ini}/M$&$E_{\rm SF}^{\rm fin}/M$&$E_{\rm SF}^{\rm fin}/E_{\rm SF}^{\rm ini}$&$M_{\rm{irr}}^{\rm fin}/M$&$Q_{\rm{BH}}^{\rm{fin}}/Q$&$Q_{\rm{SF}}^{\rm{fin}}/Q$\\
\hline
0.8&4.8E02&0.277&0.278&3.00E-05&1.32E-01&4.40E03&0.728&45 \%&55 \%\\
5.0&1.1E02&0.642&0.642&4.31E-05&3.93E-02&9.12E02&0.875&6.0 \% & 94 \% \\
20.0&4.8E01&1.756&1.757&3.13E-04&1.31E-02&4.19E01&0.924&1.0 \%& 99 \%\\
40.0&2.9E01&3.130&3.129&8.95E-04&8.02E-03&8.96E00&0.942&0.1 \% & 99.9 \%\\
\end{tabular}
\end{ruledtabular}
\end{table*}

An opposite trend is observed for the charge, as exhibited in the last two columns of Table~\ref{tab:mod1} and the bottom panel of Fig.~\ref{fg:SF2}. This figure shows a perfect charge exchange between the BH and the SF. Furthermore, the final charge in the scalar field (BH) increases (decreases) with increasing $qM$, in agreement with the last two columns of Table~\ref{tab:mod1}. Thus the charge extraction is more efficient for higher charge coupling. This observation, together with the remarks on the energy, are consistent with the computation of the irreducible mass, shown in the eighth column of the table, where one observes that $M_{\rm irr}^{\rm fin}$ approaches $M$ as $qM$ grows.

\begin{figure}[t]
\begin{minipage}{1\linewidth}
\includegraphics[width=1.0\textwidth, height=0.25\textheight]{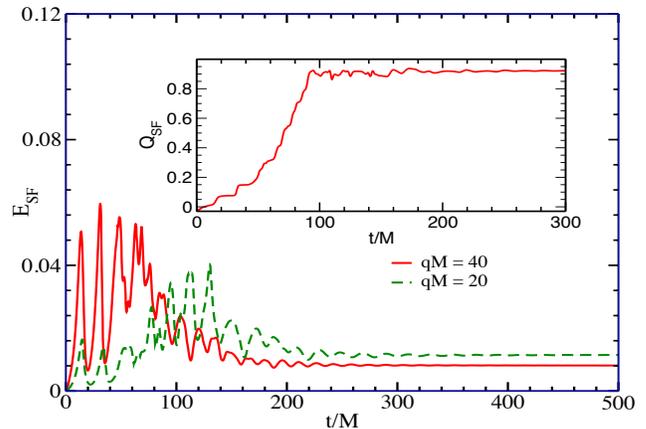} 
\caption{Bosenova the $qM =20, 40$ models. The extracted energy overshoots the final equilibrium value, and strong oscillations follow. The inset shows the SF charge for $qM=40$.}
\label{fg:SF4}
\end{minipage}
\end{figure}

{\bf {\em Bosenova.}}  The superradiant growth phase for  $qM=20,40$ is detailed in Fig.~\ref{fg:SF4}.  Whereas for models with small enough electric charge (up to $qM\sim 10$), the equilibrium phase is reached under a monotonic trend of energy extraction, for larger values of $qM$ the energy extracted clearly overshoots the final equilibrium value.  Strong oscillations of the SF energy follow, before they get damped and the system relaxes to the equilibrium phase.  In this process, some of the extracted energy is pushed back into the BH. But the charge extraction is never reversed (Fig.~\ref{fg:SF4} -- inset). This agitated and reversed (relatively steady) behaviour of the SF energy (charge), mimics that described in \cite{Yoshino:2012kn, Yoshino:2015nsa} for the energy  (angular momentum) of a test, but non-linear, SF on the Kerr background, where it was argued that it is an explosion  of the amplified SF -- akin to a bosenova -- that pushes some energy back to the BH. A more detailed analysis of this phenomenon will appear somewhere else, but we show in the supplemental material that changing the values of $\mu$ and $r_m$ does not change, qualitatively, the results above.

{\bf {\em Implications.}} We have reported the first fully non-linear evolution of a BH bomb. Our numerical simulations establish, dynamically,  that the final state of the superradiant instability, in our setup, is a hairy BH: a charged horizon surrounded by a scalar field condensate, whose real and imaginary parts oscillate, with opposite phases, at the critical frequency determined by the horizon electric potential. Together with the frequency domain perturbation analysis of~\cite{Dolan:2015dha}, our results demonstrate that these BHs are stable against superradiance, despite having  $w_{\rm c}\neq 0$, $i.e.$ non-zero horizon charge. Thus, for these hairy BHs, perturbations with $w<w_{\rm c}$ of the \textit{same  bosonic field} that constitutes the background hair, are not unstable modes.

These  hairy BHs may be considered as the charged counterparts of the hairy rotating solutions found in \cite{Herdeiro:2014goa,Herdeiro:2015gia}. The major difference between the mirror imposed here and the mass term therein is that the latter is only reflective for $w<\mu$. Thus, if there are sufficiently low frequency modes (which are the ones amplified by superradience anyway) these are gravitationally trapped and the mirror is a good model for the mass term. A further parallelism between the two cases is the bosenova-like explosion exhibited here and the one discussed for a non-linear field on the Kerr background. This supports the proposal that such rotating hairy BHs play a decisive role in the non-linear development of the rotating BH bomb in asymptotically flat spacetimes, either as long-lived intermediate states or as end-points. (Dis)proving it is an outstanding open question (see also~\cite{Bosch:2016vcp}).


\bigskip

{\bf {\em Acknowledgements.}} This work has been supported by the Spanish MINECO (AYA2013-40979-P), 
by the Generalitat Valenciana (PROMETEOII-2014-069), by the 
CONACyT-M\'exico, by the Max-Planck-Institut f{\"u}r Astrophysik, by the FCT (Portugal) IF programme, by the CIDMA (FCT) strategic project UID/MAT/04106/2013 and by the EU grants  NRHEP--295189-FP7-PEOPLE-2011-IRSES and H2020-MSCA-RISE-2015 Grant No. StronGrHEP-690904. Computations have been 
performed at the Servei d'Inform\`atica de la Universitat de Val\`encia.

\section{Supplemental Material}

In the following we provide Supplemental Material to the published paper in \textit{Physical Review Letters}. This material covers issues that, albeit not immediately relevant for understanding the paper \textit{per se}, may arise when considering the letter in more detail.  Most of this supplemental material was motivated from questions and comments by the anonymous referees, whose thorough and constructive analysis of our paper we sincerely thank.

\bigskip

{\bf {\em Initial data and constraints violation.}}
In our setup, the initial data do not
satisfy the constraints. However, since the initial data we employ in
our simulations are used in
the test-field approximation, the fact that the constraints are not
satisfied does not introduce
significant errors in the simulations. We have analysed the constraint
violations carefully and the main
features observed are
illustrated by Figs.~\ref{convergence1} and~\ref{convergence2}. These figures plot in log scale the radial
profiles of the Hamiltonian
constraint at selected times during the evolutions.
\begin{description}
\item[i)] As we noted in~\cite{Sanchis-Gual:2014nha}, where we
used the same code as
here, the larger violations of the Hamiltonian constraint take place due
to the finite-differencing
of our PDEs close to the puncture.
\item[ii)] In the absence of scalar field the violations of the Hamiltonian
constraint, dominated by the
puncture, are at a level smaller than 10$^{-5}$ - Fig.~\ref{convergence1};
\begin{figure}[h!]
\begin{center}
\includegraphics[width=0.52\textwidth]{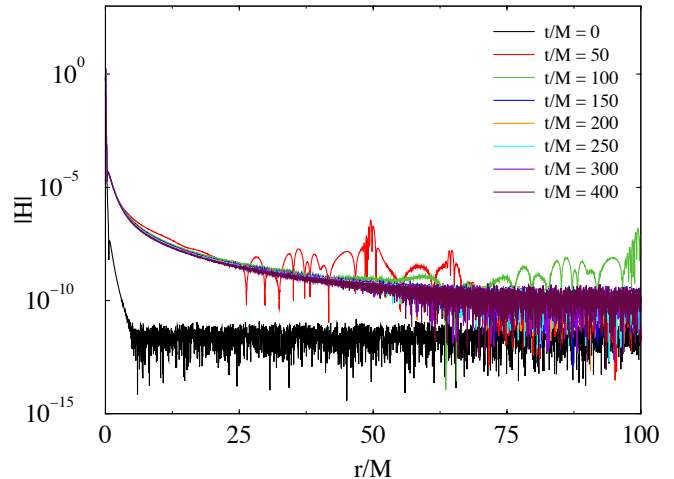}\vspace{-0.5cm}\\
\caption{Hamiltonian constraint violations for different times without scalar field ($i.e$ a Reissner-Nordstr\"om black hole with $M=1$, $Q=0.9M$).}
\label{convergence1}
\end{center}
\end{figure}
\item[iii)] In the presence of the scalar  field, as illustrated by a couple of
examples in Fig.~\ref{convergence2}, the violations
of the Hamiltonian constraint, also dominated by the puncture, albeit
slightly higher, are damped
when reaching the  final hairy black hole and settle at what is, we
believe, an acceptable value, still
of order 10$^{-5}$ outside the horizon.
\begin{figure}[h!]
\begin{center}
\includegraphics[width=0.52\textwidth]{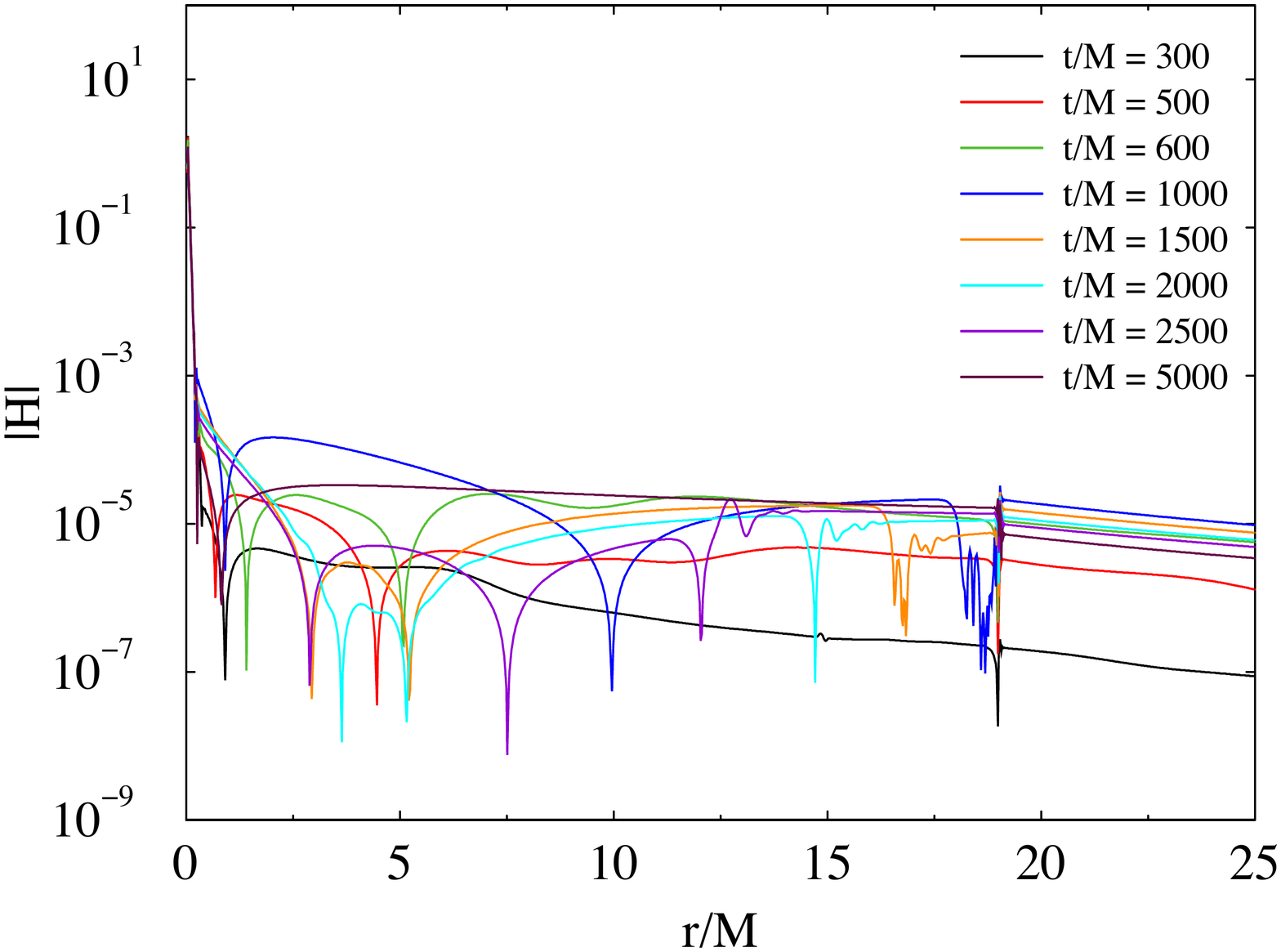}\vspace{-0.5cm}\\
\includegraphics[width=0.52\textwidth]{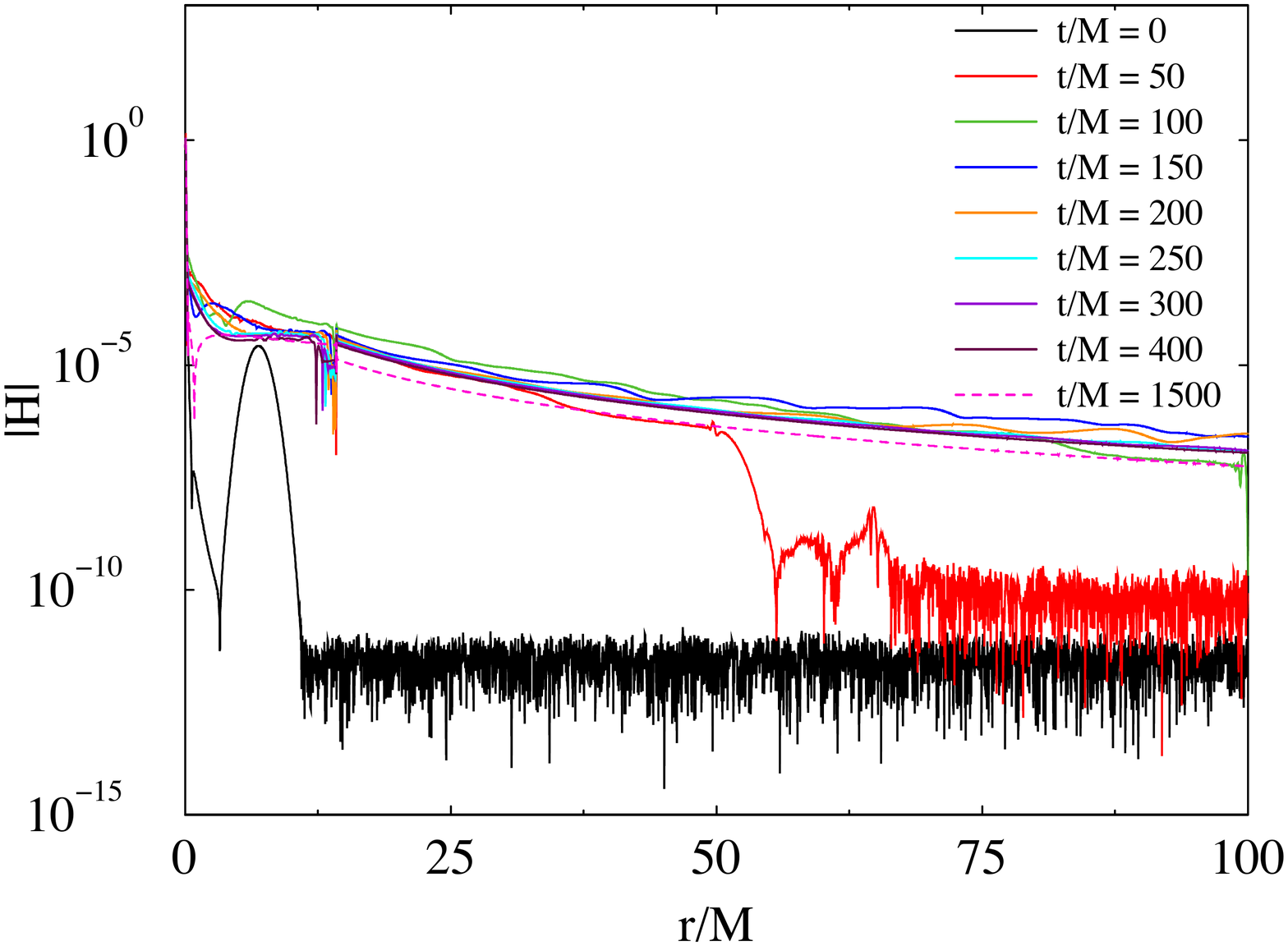}\vspace{-0.5cm}\\
\caption{Evolution of the Hamiltonian constraint violations, for $qM=5$, $r_m=19M$ (top panel) and $qM=40$, $r_m=14.2M$ (bottom panel).}
\label{convergence2}
\end{center}
\end{figure}
\item[iv)] We have observed that by increasing the resolution, the constraints
violation outside the apparent
horizon decreases at the rate given by the corresponding order of our
numerical scheme. That is, the
violation not only is small but it also converges away at the expected
second-order rate (see next section), unaffected
by the error in the vicinity of the puncture.
\end{description}
From these observations we are confident that, albeit the initial data is constraint violating, it is never -- nor initially neither during the evolution --  at unacceptable values.

%

\begin{figure}[h!]
\begin{center}
\includegraphics[width=0.52\textwidth]{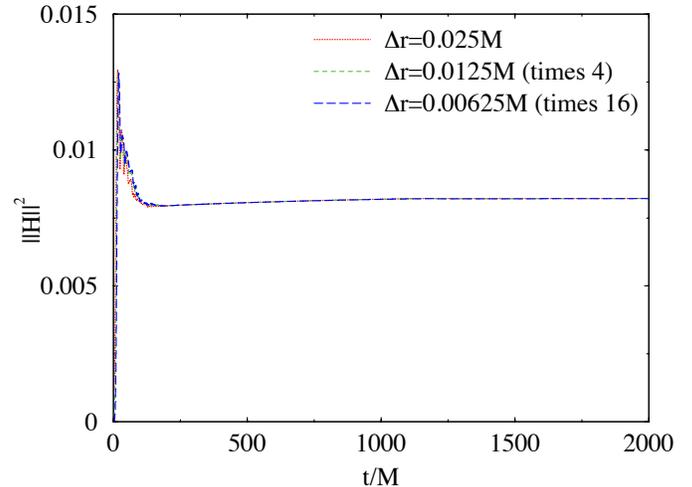}\vspace{-0.5cm}\\
\caption{Time evolution of the L2 norm of the Hamiltonian constraint for an initial scalar field pulse with $qM=40$ and $r_m=14.2M$. The plot shows results for three different resolutions, rescaled by the factors corresponding to second-order convergence.}
\label{convergence3}
\end{center}
\end{figure}

{\bf {\em Convergence.}} In our simulations, we observe second order convergence, as expected for this code and as illustrated in Fig.~\ref{convergence3}.

\begin{figure}[h!]
\begin{center}
\includegraphics[width=0.52\textwidth]{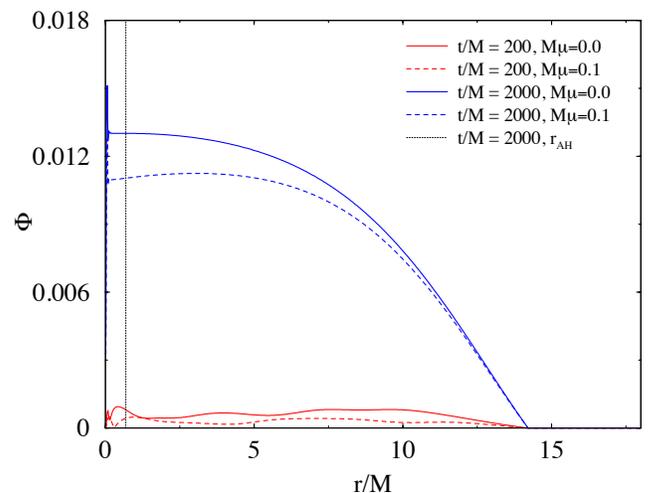}\vspace{-0.5cm}\\
\caption{Scalar field magnitude at two different time slices, for two different values of the scalar field mass, in terms of the radial coordinate, for $qM=5$ and $r_m=14.2M$. The vertical line marks the location of the apparent horizon at the final time.}
\label{mass}
\end{center}
\end{figure}

\begin{figure}[h!]
\begin{center}
\includegraphics[width=0.52\textwidth]{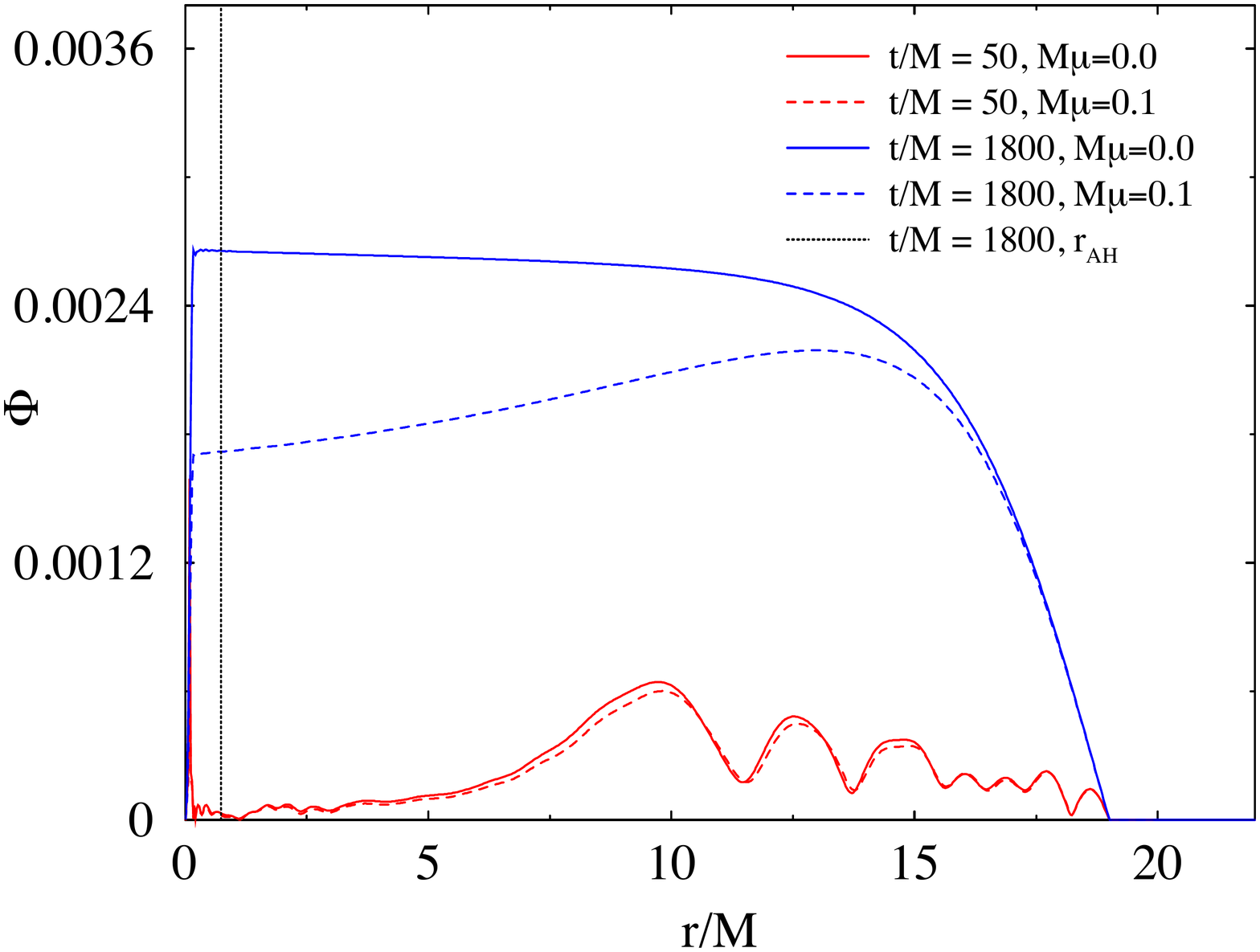}\vspace{-0.5cm}\\
\caption{Scalar field magnitude at two different time slices, for two different values of the scalar field mass, in terms of the radial coordinate, for $qM=40$ and $r_m=19M$. The vertical line marks the location of the apparent horizon at the final time.}
\label{mass2}
\end{center}
\end{figure}

{\bf {\em Sensitivity of the results to $\mu$.}} The choice to study a non-zero scalar field mass was ``conceptual" since there are no known charged particle with zero mass in the standard model of particle physics. Thus we considered it would be more natural to take a non-zero mass to study our charged scalar field. In any case we have also performed simulations with zero mass as well, and some comparative results are presented in Figs.~\ref{mass} and~\ref{mass2}. As can be seen there is no qualitative difference in the massless case: a hairy black hole still forms with a scalar field profile. The major difference is the one already anticipated in the manuscript: in the massless case the scalar field profile has no maximum outside the apparent horizon, in agreement with the results of~\cite{Dolan:2015dha}.

\begin{figure}[h!]
\begin{center}
\includegraphics[width=0.52\textwidth]{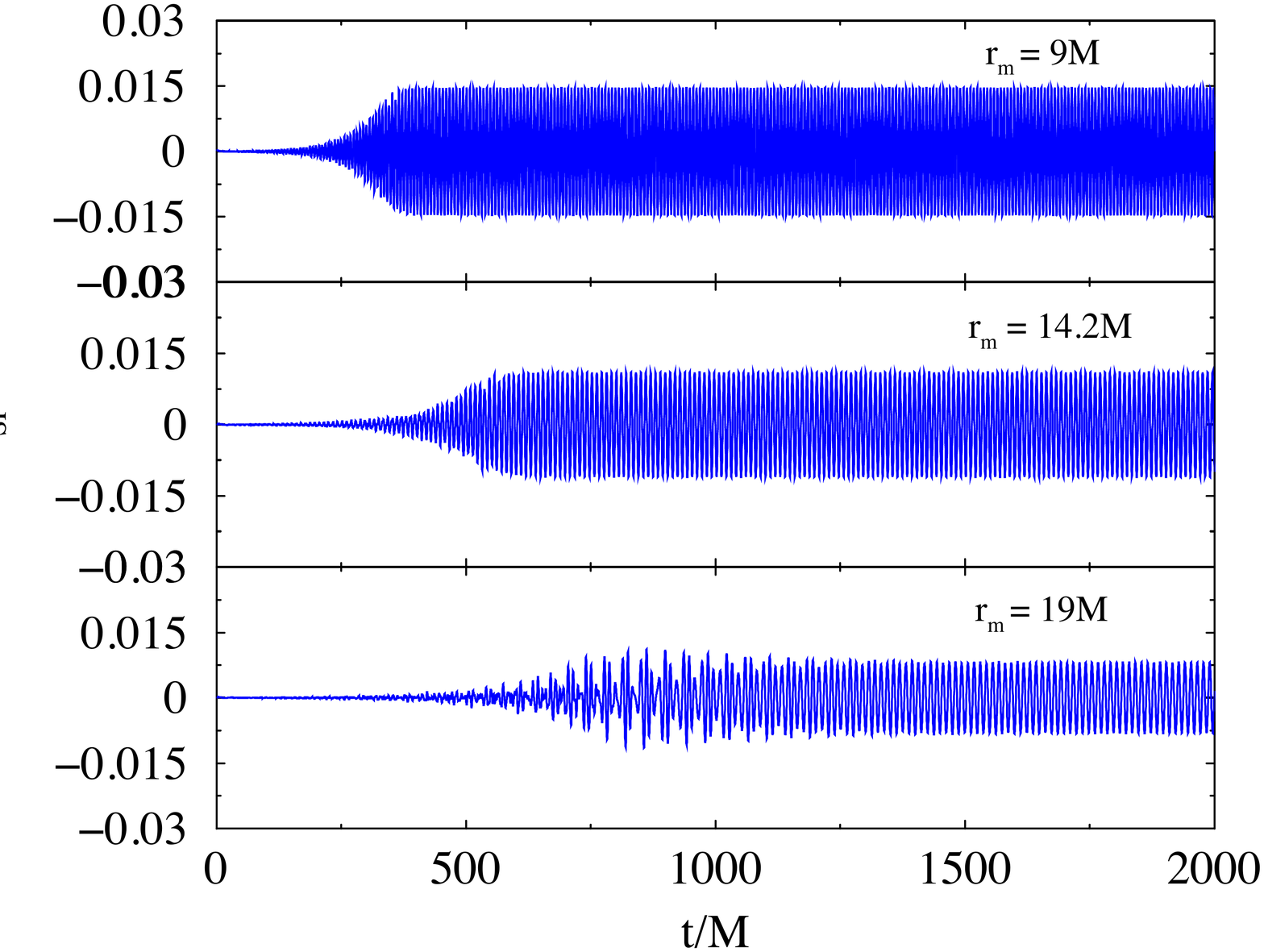}
\caption{Plots equivalent to Fig. 1 in the manuscript, for $qM=5$, varying $r_m$.}
\label{mirror}
\end{center}
\end{figure}

\begin{figure}[h!]
\begin{center}
\includegraphics[width=0.52\textwidth]{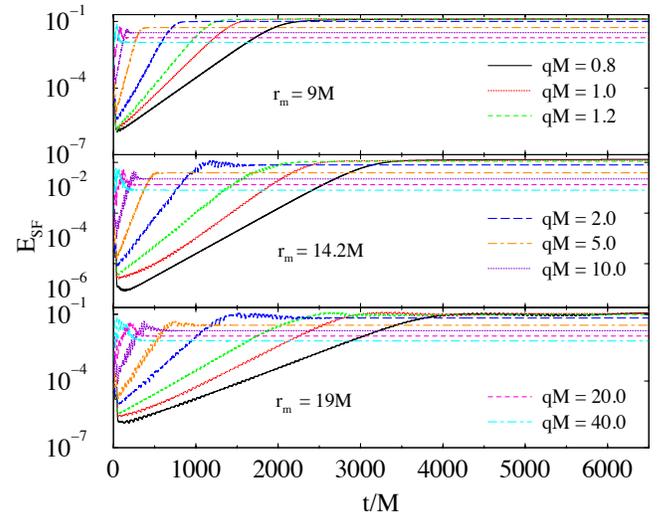}
\caption{Plots equivalent to Fig. 3 (top panel)  in the manuscript, varying $r_m$.}
\label{mirror2}
\end{center}
\end{figure}

\begin{figure}[h!]
\begin{center}
\includegraphics[width=0.52\textwidth]{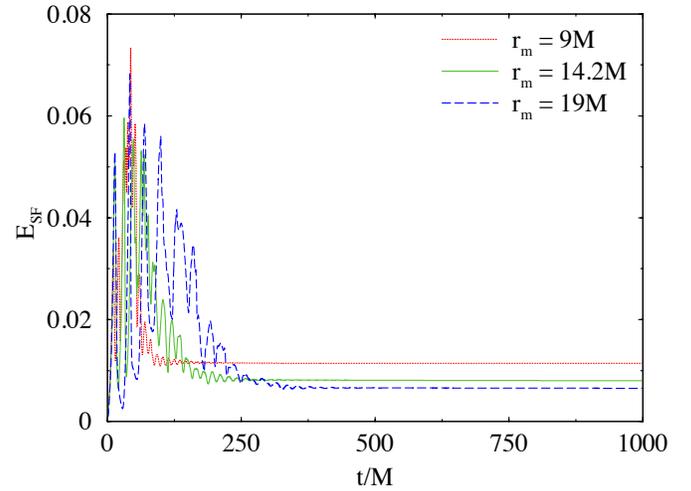}
\caption{Plots equivalent to Fig. 4 (main panel) in the manuscript, varying $r_m$ ($qM=40$).}
\label{mirror3}
\end{center}
\end{figure}

{\bf {\em Sensitivity of the results to $r_m$.}} We have studied different values of $r_m$ with similar results to the ones presented in the letter. The value chosen in the letter is illustrative. In Fig.~\ref{mirror}, Fig.~\ref{mirror2} and Fig.~\ref{mirror3}  we exhibit  plots equivalent to Fig. 1, Fig. 3 (top panel) and Fig. 4 (main panel) in the letter, respectively, varying $r_m$. As can be observed the results are qualitatively similar.

\bibliography{num-rel}


\end{document}